\documentclass[aps,twocolumn]{revtex4-1}
\usepackage{rotating}
\usepackage{verbatim}
\usepackage[normalem]{ulem}
\usepackage{bm} 
\usepackage{amssymb,amsmath,
amsthm,epsfig,subfigure}  
  
 \makeatletter  \newcommand{\Rmnum}[1]{\expandafter\@slowromancap\romannumeral #1@}

\newcommand{\be}{\begin{equation}}
\newcommand{\ee}{\end{equation}}
\newcommand{\ber}{\begin{eqnarray}}
\newcommand{\eer}{\end{eqnarray}}
\usepackage[belowskip=-30pt]{caption}
\makeatother

\usepackage{color}
\usepackage{bm} \usepackage{amssymb,amsmath,amsthm,epsfig,subfigure}

\begin{document}

\title{Time-dependent Probability Density Functions and Information Geometry \\
in the
Fusion L-H Transition}
\author{Eun-jin Kim$^{1,2}$ and Rainer Hollerbach$^3$}
\affiliation{
$^1$ Fluid and Complex Systems Research Centre, Coventry University, Coventry
CV1 2TT, UK\\
$^2$ School of Mathematics and Statistics, University of Sheffield, Sheffield S3 7RH, UK\\
$^3$ Department of Applied Mathematics, University of Leeds, Leeds LS2 9JT, UK
}
\vspace{1cm}
\begin{abstract}
We report a first study of time-dependent Probability Density Functions (PDFs) in the Low-to-High confinement mode (L-H) transition by extending the previous prey-predator-type model (Kim \& Diamond, Phys.\ Rev.\ Lett.\ 91, 185006, 2003) to a stochastic model. We 
{highlight the limited utility of mean value and variance in understanding the L-H transition by showing} strongly non-Gaussian PDFs, with the number of peaks changing in time. We also propose a new information geometric method by using information length, dynamical time scale, and information phase portrait, and show their utility in forecasting transitions and self-regulation between turbulence and zonal flows.
{In particular, we demonstrate the importance of intermittency (rare events of large amplitude) of zonal flows that can play an important role in promoting the L-H transition.} 
\end{abstract}

\maketitle
\section{Introduction} 
An important example of non-equilibrium processes is found in magnetically confined fusion plasmas (ionized gas) which aim to achieve a controlled generation of energy, mimicking nuclear reactions naturally taking place in the Sun and stars. The key challenge in fusion has been proper confinement of hot plasmas with a temperature greater than $10^7$ $^\circ$K (hotter than the center of the Sun!) inside the device, which itself is at most at room temperature. This large temperature difference across a few-meter wide device is very unstable, causing turbulent (anomalous) transport and thus confinement degradation, or even the termination of fusion operation.\\

The L-H (Low-to-High) transition, first discovered in 1980s, marked one of the greatest discoveries in fusion research \cite{ASDEX} where plasma confinement improved dramatically when an input power exceeded a critical value. This constitutes an intriguing example of self-organization \cite{zan,diamond05,Chang2017,Schmitz2017,kim03,Malkov,Miki,DMK13,zhu,bian} where plasmas organize themselves into an ordered, High-confinement (H) mode from a Low-confinement (L) mode triggered by the formation of large-scale shear (mean, zonal) flows which reduce turbulent transport \cite{zonalflow,hahm,shat,gb,KIM06}. While being reproduced in different devices, the realization of the H-mode for future burning plasmas (e.g.\ \$20 billion ITER project) remains a critical issue \cite{Schmitz2017,yasmin2019}, with controversial issues including threshold power scaling, the effects of density, magnetic geometry and neutrals, triggering mechanisms and causality relations, hysteresis, etc.\ \cite{yasmin2019}. This has far reaching implications for other self-regulating systems in nature (e.g. astro/geophysical, atmospheric sciences, etc.).
Previously employed statistical methods include moments (mean value, variance), spectral/wavelet analysis, bicoherence, phase relation, turbulence-flow energy transfer, fluxes, transport coefficients, etc.\\

This paper reports the first study of time-dependent Probability Density Functions (PDFs) \cite{PDF16} in the L-H transition, which are invaluable to understand strongly time-dependent fluctuations (intermittency) \cite{phasetransition}, often associated with transitions. The latter leads to non-Gaussian, non-stationary PDFs \cite{intermittency,phasetransition}, with the limited validity of mean value/variance, or stationary PDFs. Time-dependent PDFs also enable us to understand the correlation/causality and hysteresis from the perspective of information theory. In simple terms, instead of the physical variables themselves, we consider statistical states of different variables and how they change in time and are correlated with each other. Here, the changes in ``statistical states'' are quantified by dimensionless numbers from time-dependent PDFs that are {\it invariant under (time-independent) change of variables}, which can be directly compared with each other unlike physical variables having different units. Specifically, we quantify how each variable passes through statistically different states during the evolution (see below). Our proposed method captures the dynamics rather than stationary properties of the L-H transition, which we believe to be crucial since the latter is a dynamical process, evolving over time. \\

{\it Information length.} We begin by summarizing how to calculate the change in statistical states for a stochastic variable $x$ which has a time-dependent PDF $p(x,t)$. By calculating an infinitesimal relative entropy between $p(x,t)$ and $p(x,t +\delta t)$ as $ \delta t \to 0$, and then summing the square root of the infinitesimal relative entropy along
the path, we define the (dimensionless) information length ${\cal L}(t)$ \cite{entropy18,HK16,KIM16,KH17,phasetransition,JSTAT,HM2019,HK19,KIM20}
\begin{eqnarray}
{\cal{L}} (t) = \int_0^{t} \frac{dt_1}{\tau(t_1)},&&\,\quad
\frac{1}{\tau(t)^2}  = \int dx \frac {1} {p(x,t)}
 \left [\frac {\partial p(x,t)} {\partial t} \right]^2.\quad
 \label{eq01}
 \end{eqnarray}
The unit of $\tau$ in Eq.~(\ref{eq01}) is time, representing a dynamical time unit for information change; ${\cal L}(t)$ then measures the clock time in units of $\tau(t)$, and quantifies the total number of statistically different states that $x$ passes through between time $0$ and $t$, starting from some initial PDF $p(x,0)$. {In simple terms, ${\cal L}(t)$ quantifies the cumulative change in $p(x,t)$ taking into account the uncertainty due to a finite width of $p(x,t)$.} Unlike more traditional methodologies such as entropy, relative entropy, Jensen divergence, etc., ${\cal L}(t)$ depends on $p(x,t')$ for all $ t' \in [0,t]$ and is thus a path-dependent quantity. This path-dependence is ideally suited for understanding a long memory and hysteresis involved in phase transitions \cite{phasetransition} such as the L-H transition. It can also be used to quantify attractor structures in relaxation processes \cite{entropy18,HK16,KH17}, providing an alternative to a Lyapunov exponent to characterize chaos. {A strong correlation between two switching species was captured by the same evolution of ${\cal L}(t)$ of these two species \cite{HK19}.}\\

For a system with $m$ variables $x_i$ ($i=1,2,..m$), we can extend Eq.~(\ref{eq01}) to
\begin{eqnarray}
&& {\cal{L}}_{x_i} (t) = \int_0^{t} \frac{dt_1}{\tau_{x_i}(t_1)},
\label{eq002}\\
&& \frac{1}{[\tau_{x_i}(t)]^2}  = \int dx_i \frac {1} {p(x_i,t)}
 \left [\frac {\partial p(x_i,t)} {\partial t} \right]^2 \equiv{\cal{E}}_{{x_{i}}},\,\,\,
 \label{eq001}
 \end{eqnarray}
where  $p(x_i,t) = \int \Pi_{j \ne i}  (dx_j) \, p(x_1, x_2,..., x_m)$ is a marginal PDF of $x_i$. Note that $\tau_{x_i}$ and ${\cal L}_{x_i}$ depend on the path of $x_i$, and the correlation or causality among different variables can be inferred by comparing $\tau_{x_i}$ for different $x_i$, as demonstrated below. \\

The remainder of this paper is organised as follows. Section II provides our stochastic model. The corresponding Fokker-Planck equation is solved numerically in Section III. Sections IV and V present results and conclusions, respectively.

\section{Stochastic model} 
We apply Eqs.~(\ref{eq002})-(\ref{eq001}) to the stochastic version of the previous prey-predator model of the L-H transition \cite{kim03}. Specifically, despite highly nonlinear multiscale interactions involved in the L-H transition, the very nature of self-organization (universality and robustness) \cite{JSTAT,DAM2017} makes it possible to capture qualitative behaviour of the L-H transition through reduced models and to explore different parameters at a low cost \cite{Malkov,Miki,DMK13,zhu,bian}. In \cite{kim03}, turbulence amplitude $\epsilon$, zonal flow $v$ and density gradient $N$ are governed by
\begin{eqnarray}
\frac{\partial \epsilon}{\partial t} &= & N \epsilon - a_1\epsilon^2
 - a_2 V^2 \epsilon - a_3 v^2 \epsilon,
\label{eq1}\\
\frac{\partial v}{\partial t} &= &  \frac{b_1 \epsilon v}{1 + b_2 {V}^2} -b_3 v,
\label{eq2}\\
\frac{\partial N}{\partial t} &= & - c_1 \epsilon N - c_2 N + Q.
\label{eq3}
\end{eqnarray}
Here $a_{i}$, $b_{i}$ and $c_{i}$ are non-negative constants, ${V}=d N^2$ (with $d$ a positive constant) is the mean flow, and $Q$ is the external heating that ultimately drives the entire system. {Eqs.\ (\ref{eq1})-(\ref{eq3}) are identical to 
Eqs.\ (6)-(8) in \cite{kim03}, $v$, $\epsilon$ and $N$ here corresponding to $V_{ZF}$, ${\cal E}$ and ${\cal N}$ in \cite{kim03}.}\\

The right side of Eq.~(\ref{eq1}) represents the linear growth of turbulence by the density gradient and turbulence damping due to turbulence nonlinear interaction, mean flow and zonal flow, respectively. Eq.~(\ref{eq2}) similarly represents the zonal flow growth from turbulence, subject to the mean flow damping ($1+ b_2{V}^2$), and linear (collisional) damping. Eq.~(\ref{eq3}) represents the damping of the density gradient due to turbulence and neo-classical/collisional effect, and the density gradient growth due to the external heating $Q$. Eqs.~(\ref{eq1})-(\ref{eq3}) support the L-H transition either with or without going through limit-cycle oscillation (I-phase), depending on precise parameter values and $Q$. {This I-phase is due to the self-regulation between $v$ and $\epsilon$; for sufficiently large $Q$ the dithering phase enters a quiescent H-mode where $v=\epsilon=0$ \cite{kim03,Malkov,DMK13,DAM2013}. In this model, zonal flows trigger the transition to a quiescent H mode by lowering the power threshold, while mean flow $V$ locks the plasma in the H-mode.} It is not our intention here to explore all possible cases, but to focus on a limited set of calculations to focus attention on the effect of stochasticity and new methods. Similarly, detailed bifurcation analyses can be done \cite{Malkov,DAM2013}, but would be of limited interest for the time-dependent $Q(t)$ that we consider here. 
{Fluctuating (oscillatory) $Q$ was shown to help the L-H transition by lowering the constant part of the power threshold \cite{DMK13}. We will show a similar effect of stochasticity in $\epsilon$ and $v$.} \\

For a stochastic model, it turns out to be better to work with $x=\pm \sqrt{\epsilon}$. 
Solving the Fokker-Planck equation (\ref{eq8}) below for $x$ instead of $\epsilon=x^2$ allows us to avoid the need to impose the `boundary' $\epsilon \to0$ and instead to deal with the much more natural boundaries $x\to\pm\infty$. This also makes additive noise more straightforward than it would be in the original $\epsilon$ formulation. One further simplification to facilitate the numerical calculation of PDFs via Eq.\ (\ref{eq8}) is to assume that $N$ evolves sufficiently rapidly to approximate Eq.~(\ref{eq3}) as 
\begin{equation}
N=\frac{Q}{ c_1 x^2 +c_2}.
\label{eq4}
\end{equation}
{
Eqs.\ (\ref{eq1})-(\ref{eq2}) and (\ref{eq4}) were also proposed as a reduced L-H transition model in \cite{DAM2013}, and the even more drastic approximation $N={Q}/{c_2}$ was used in \cite{bian} to investigate the effect of intermittency, while mean flow was neglected completely in \cite{MD11} to understand bistability of zonal flows and geodesic acoustic modes. Numerical solutions of either the original set of three or the reduced set of two ODEs yield qualitatively the same results, but for the corresponding Fokker-Planck equation the reduction from three to two variables results in substantial computational savings, as discussed below.}\\

By introducing two independent $\delta$-correlated Gaussian stochastic noises $\xi$ and $\eta$ {in Eqs (\ref{eq1}) and (\ref{eq2}) respectively} \cite{Fokker}, we formulate stochastic equivalents of Eqs.~(\ref{eq1})-(\ref{eq2}):
\begin{eqnarray}
\frac{d x }{d t} =  f + \xi,&&\,\,\,\,
f = \frac{1}{2} \Bigl[ N    - a_1 x^2  - a_2 V^2 - a_3 v^2 \Bigr] x,
\label{eq5}\\
\frac{d v}{d t} =   g + \eta,&&\,\,\,\,
g = \frac{b_1 x^2 v}{1 + b_2 V^2} -b_3 v,
\label{eq6}
\end{eqnarray}
and $N$ given by Eq.~(\ref{eq4}). The noise terms satisfy
\begin{eqnarray}
\langle \xi(t) \xi(t') \rangle= 2D_x \delta(t-t'), \,\,&&
\langle \eta(t) \eta(t') \rangle= 2D_v \delta(t-t'),
\nonumber \\
\langle \xi(t) \eta(t') \rangle=0, \,\,\,\,&& \langle \xi \rangle= \langle \eta \rangle= 0,
\label{eq7}
\end{eqnarray}
where the angular brackets denote averages. $D_x$ and $D_v$ are the amplitudes of the stochastic noise $\xi$ and $\eta$, affecting $x$ and $v$ respectively.\\

\section{Fokker-Planck equation}

The Fokker-Planck equation \cite{Fokker} for the joint PDF $p=p(x,v,t)$ corresponding to Eqs.~(\ref{eq5})-(\ref{eq7}) is
\begin{eqnarray}
\frac{\partial p}{\partial t}
&=& -\frac{\partial}{\partial v}(g\,p)  - \frac{\partial}{\partial x}(f\,p)
+ D_x \frac{\partial^2p}{\partial x^2}+ D_v \frac{\partial^2p}{\partial v^2}.
\label{eq8}
\end{eqnarray}
{In contrast, without the simplification from using Eq.\ (\ref{eq4}), the Fokker-Planck equation would describe the joint PDF $p(x,v,N,t)$ depending on three random variables $(x,v,N)$ in addition to time $t$; using Eq.\ (\ref{eq4}) thus reduces the dimensionality of the numerical problem from three to two `spatial' variables, allowing a far more thorough exploration of parameter values, as well as more narrowly peaked PDFs.}

The numerical solution of (\ref{eq8}) involves second-order finite-differencing, with grid spacings as small as $10^{-3}$ in both $x$ and $v$. The time-stepping is second-order Runge-Kutta, with time-steps as small as $2\cdot10^{-5}$. 
Taking a box size with $x_{max}=v_{max}=2$ is sufficiently large to be a good approximation to $x,v\to\infty$; that is, the total probability $\iint p\,dx\,dv$ remains conserved within $10^{-4}$ or better for all runs presented here.
\begin{figure}[b]
\begin{center}
\includegraphics[scale=0.9]{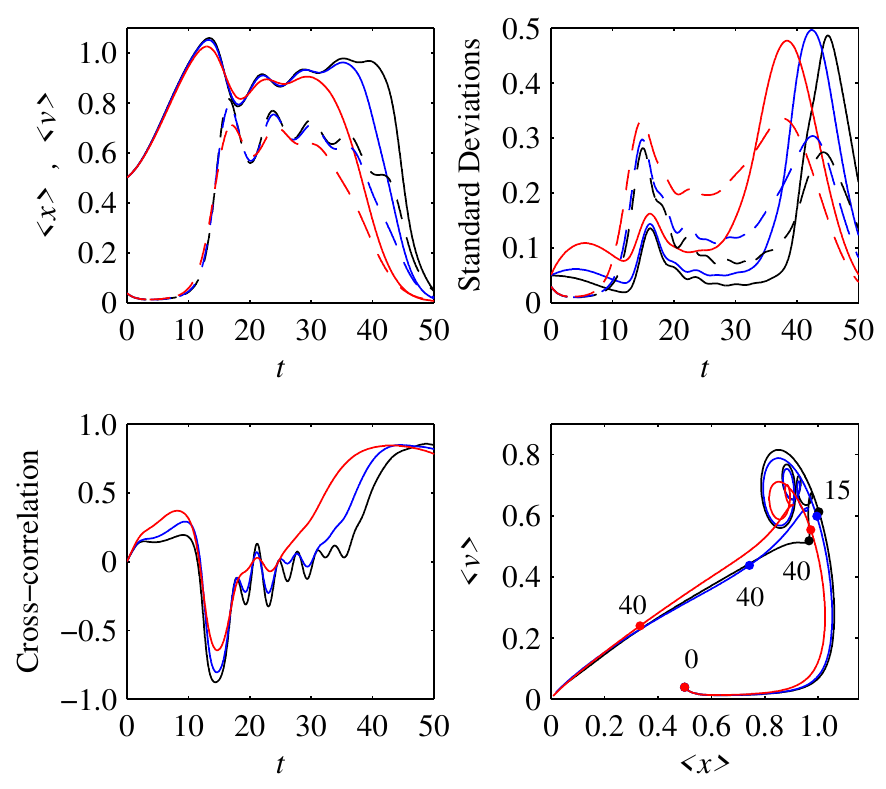}
\vspace{-0.2cm}
\caption{The first (top left) panel shows the averages $\langle x\rangle$ (solid lines) and $\langle v\rangle$ (dashed lines) against time. The second (top right) panel shows the corresponding standard deviations $\sigma_x$ (solid) and  $\sigma_v$ (dashed). The third (bottom left) panel shows the cross-correlation. The fourth (bottom right) panel is the phase portrait in $(\langle x\rangle,\langle v\rangle)$. {The dots with associated numbers correspond to $t=0, 15, 40$.} For all panels, [black,{\color{blue}blue},{\color{red}red}] correspond to $D_x=[1,4,16]\cdot10^{-4}$, respectively.}
\label{fig1}
\end{center}
\end{figure}

{In order to facilitate the comparison with the previous deterministic model \cite{kim03}, we use the same parameter values} $a_1=0.2$, $a_2=a_3=0.7$, $b_1=1.5$, $b_2=b_3=1$, $c_1=1$, $c_2=0.5$, and $d=1$ as those in \cite{kim03}.
For the input power we take $Q(t)=0.1+0.03t$, for $t\in[0,50]$, so $Q$ ramps up from 0.1 to 1.6. The initial condition is $p(x,v,0)\propto\exp[-((|x|-0.5)^2-v^2)/5\cdot10^{-3}]$. Other initial conditions with small values of $x$ and $v$ were also investigated and yield similar results. For the noise terms $D_x$ and $D_v$ we explored the range $10^{-4}$ and greater. Varying $D_v$ turned out to have relatively little impact, so we fix $D_v=10^{-4}$, and present results for $D_x=[1,4,16]\cdot10^{-4}$.
{Since the prediction from the deterministic model in \cite{kim03} has been reproduced in various laboratory experiments in terms of the time-evolution of the mean values, 
the results from our stochastic model are expected to capture experimental results qualitatively.}

From the joint PDF $p(x,v,t)$ we can also obtain the marginal PDFs $p(x,t)=\int p(x,v,t)\,dv$ and $p(v,t)=\int p(x,v,t)\,dx$, and then compute the information length diagnostics ${\cal E}_x$ and ${\cal E}_v$ ($\tau_x(t)$ and $\tau_v(t)$) as in Eq.~(\ref{eq001}), and ${\cal L}_x (t) = \int_0^t dt_1/\tau_x(t_1)$ and ${\cal L}_v (t) = \int_0^t dt_1/ \tau_v(t_1)$ as in Eq.~(\ref{eq002}).
Although other statistical quantities including entropy and Fisher information were also calculated, they were less informative in capturing the L-H transition and thus are not presented below. 

\section{Results}

\subsection{Mean, variance, phase portrait}

Figure \ref{fig1} shows the average quantities $\langle x\rangle$, $\langle v\rangle$, the standard deviations $\sigma_x=\sqrt{\langle (x-\langle x\rangle)^2\rangle}$ and $\sigma_v=\sqrt{\langle (v-\langle v\rangle)^2\rangle}$, and the (normalized) cross-correlation $\langle (x-\langle x\rangle)(v-\langle v\rangle)\rangle/(\sigma_x\sigma_v)$. Note that the average $\langle\cdot\rangle$ refers to the mean value over the first quadrant $x,v>0$ only, that is, $\langle f\rangle\equiv\int_0^2\int_0^2 f\,p\,dx\,dv$. Following the abrupt increase in  $\langle v\rangle$ at $t \approx 11$ for all $D_{x}$, the dithering I-phase starts where $\langle x\rangle $ and $\langle v\rangle$ oscillate. The dithering phase ends when $\langle x \rangle$ and $\langle v \rangle$ both collapse back towards zero, corresponding to the transition to the H-mode. \\

\begin{figure}[b]
\begin{center}
\includegraphics[scale=0.9]{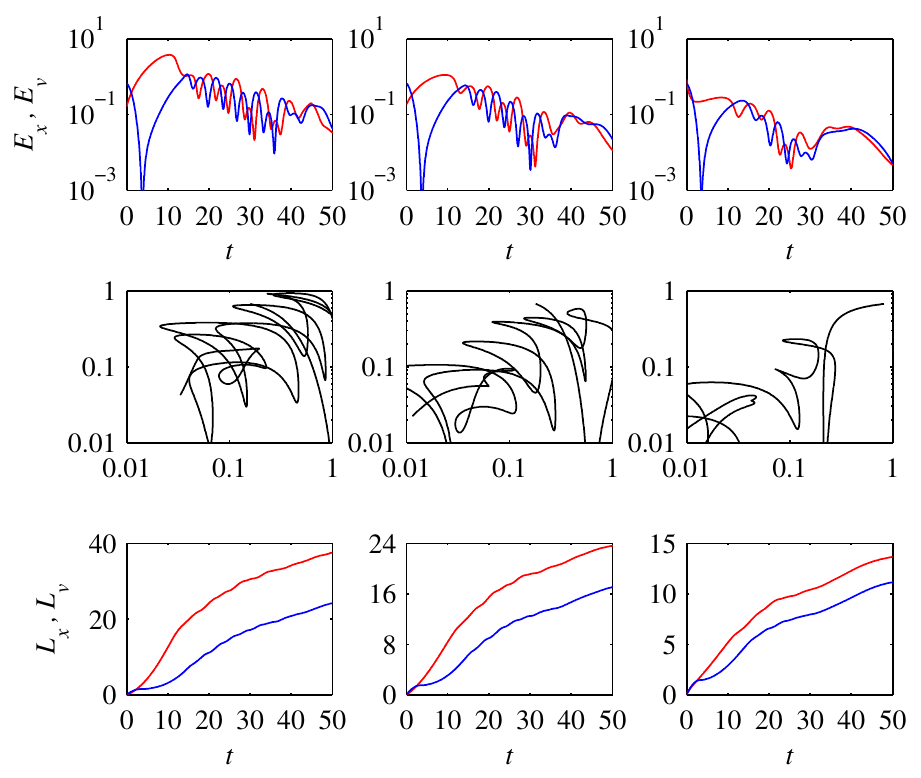}
\caption{From left to right the three panels are $D_x=[1,4,16]\cdot10^{-4}$. The top row shows ${\cal E}_x$ (red) and ${\cal E}_v$ (blue) against time. The middle row shows information phase portrait in (${\cal E}_x$, ${\cal E}_v$) plane using logarithmic scales. The bottom row shows the corresponding ${\cal L}_x$ (red) and ${\cal L}_v$ (blue).\\}
\label{fig2}
\end{center}
\end{figure}

\begin{figure*}
\begin{center}
\includegraphics[scale=0.9]{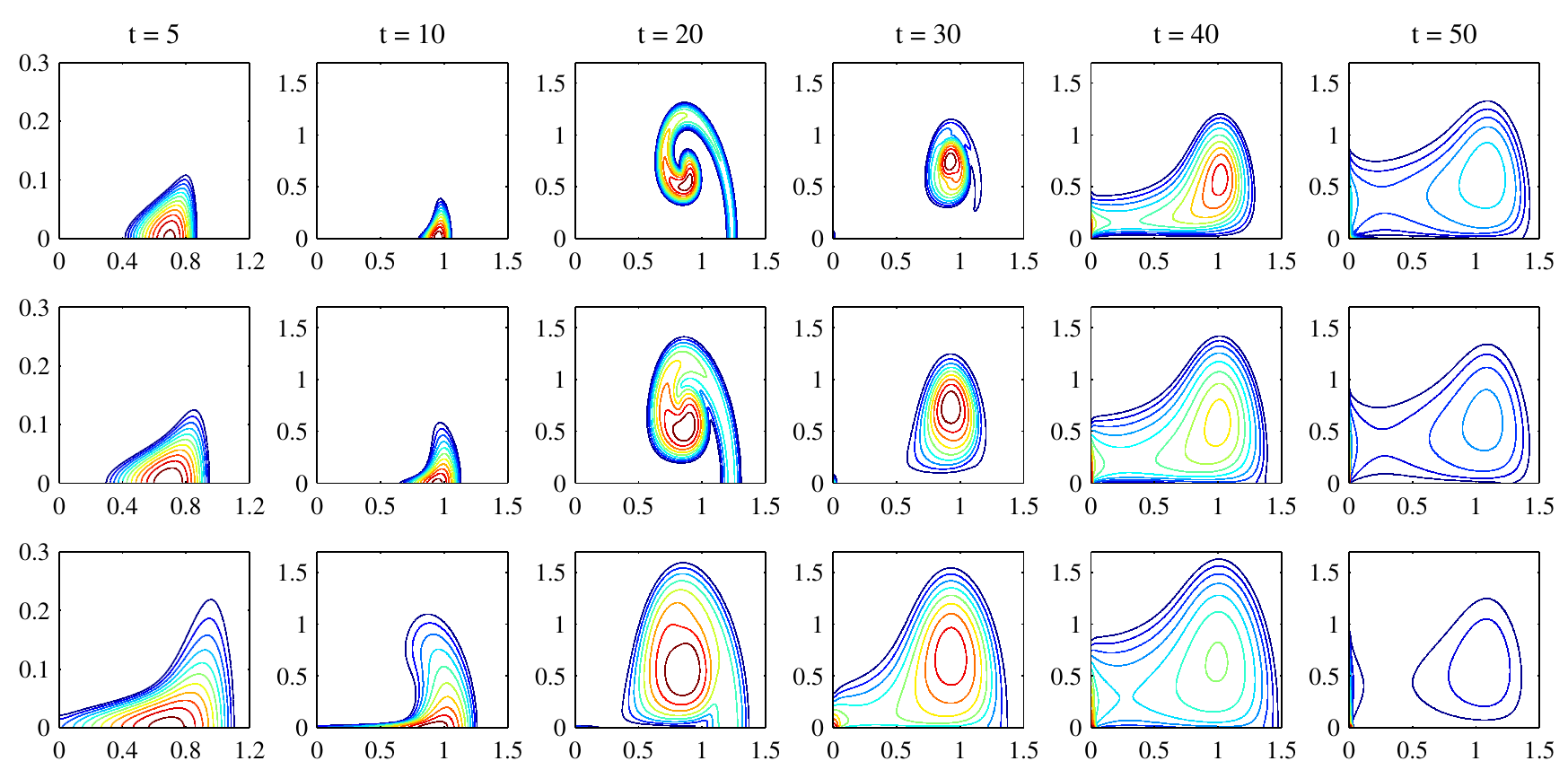}
\caption{Contour plots of the joint PDFs $p(x,v,t)$, with $x$ and $v$ on the horizontal axis and $v$ on the vertical axis. From top to bottom the three rows are $D_x=[1,4,16]\cdot10^{-4}$. The six panels are at times $t=5,10,20,30,40,50$, as also indicated at the top of the figure. Notice how $t=5$ has different $(x,v)$ ranges than the later times. Contours are on a logarithmic scale, $10^{-3}$, $10^{-2.5}$, $10^{-2}$, etc., to show the overall structure and not just the peaks.\\}
\vspace{0.5cm}
\label{fig3}
\end{center}
\end{figure*}

The self-regulation between $x$ and $v$ can be inferred from the phase shift, as the maxima of $\langle x \rangle$ preceed those of $\langle v \rangle$; note similarly the negative sign of the cross-correlation at $t\approx15$, and the following (with alternative sign) fluctuations. The large increase in $\sigma_x$ and $\sigma_v$ at the beginning and end of the dithering phase also signifies the importance of fluctuations around the transitions.
{Larger values of $D_x$ help entering the H-mode at earlier time, and thus at smaller power $Q$ ($=0.01+0.03t$). That is, greater stochastic noise helps the L-H transition by lowering power threshold. Also, the upper left panel of Figure \ref{fig1} reveals that the larger $D_x$, the smaller the maximum value of $\langle v \rangle$ and $\langle x \rangle$ around $t \approx 15$. However, exactly the opposite tendency is observed in $\sigma_x$ and $\sigma_v$ in the upper right panel of figure \ref{fig2} where the larger maximum values of $
\sigma_x$ and $\sigma_v$ occur for a larger $D_x$. Their overall effect can only be understood by investigating PDFs, and will be discussed below when discussing Figures~\ref{fig4}-\ref{fig5}.}

\subsection{Information length diagnostics}

Figure \ref{fig2} shows the information length diagnostics ${\cal E}_x$, ${\cal E}_v$, ${\cal L}_x$, and ${\cal L}_v$. Plotted as functions of time, ${\cal E}_x$ and ${\cal E}_v$ exhibit an intricate series of oscillations in the I-phase, with similar magnitudes overall but alternating in which is larger (see below for detailed discussion). When ${\cal E}_x$ and ${\cal E}_v$ cross, the time scales of $p(x,t)$ and $p(v,t)$ match (reminiscent of resonance), implying a strong correlation between the two. Furthermore, right before the transition to the I-phase we have ${\cal E}_x\gg{\cal E}_v$, corresponding to $\tau_x \ll \tau_v$, which suggests that $x$ (turbulence) is leading the dynamics. Also, larger $D_x$ not only shortens the extent of the dithering phase, but further dampens out these oscillations, resulting in a significantly reduced number of crossings.\\

By comparing ${\cal E}_{x}$ and ${\cal E}_{v}$ in figure \ref{fig2} with $\langle x \rangle$ and $\langle v \rangle$ in figure \ref{fig1}, we make the following important observations. First, ${\cal E}_v$ starts increasing at much earlier time (e.g. $t \approx 3.9$ for $D_x =10^{-4}$, $t \approx 3.7$ for $D_x =16 \cdot 10^{-4}$) than $\langle v \rangle$ does (at $t \approx 11$). The maximum in ${\cal E}_x$ occurs at earlier times (e.g. $t \approx 10.5$ for $D_x = 10^{-4}$, $t \approx 9.5$ for $D_x = 16 \cdot 10^{-4}$) than $\langle x \rangle$ (at $t \approx 13.5$). These results suggest that ${\cal E}_x$ and ${\cal E}_v$ forecast the transition to the I-phase earlier (better) than mean values. 
Third, the effect of $D_x$ is more pronounced in ${\cal E}_x$ and ${\cal E}_v$ than in $\langle x\rangle$ and $\langle v\rangle$. For instance, the maximum values of ${\cal E}_x$ are $\approx$ 4 to 0.6 for $D_x = [1,16] \cdot 10^{-4}$ while that of $\langle x \rangle$ is approximately the same, reflecting the sensitivity of our diagnostics.
{Fourth, the transition to H-mode can also be inferred from the loss of the 
(fast) oscillation around $\tau_x=\tau_v$ as well as the sudden increase
in ${\cal E}_x$ and ${\cal E}_v$ due to the loss of self-regulation.} \\

As noted above, the oscillations between ${\cal E}_x$ and ${\cal E}_v$ during dithering manifest the competition between turbulence and zonal flows as a result of self-regulation. To visualize this, we show the information phase portrait of ${\cal E}_x$ against ${\cal E}_v$ in the middle row in figure \ref{fig2}
where ${\cal E}_x$ and ${\cal E}_v$ oscillate around a straight line ${\cal E}_x = {\cal E}_v$ ($\tau_x = \tau_v$).\\

Finally, the bottom row of figure \ref{fig2} shows the information length ${\cal L}_x$ and ${\cal L}_v$.  Note that the slope of ${\cal L}_x$ and ${\cal L}_v$ are $\sqrt{{\cal E}_x}$ and $\sqrt{{\cal E}_v}$, respectively. Overall, ${\cal L}_x$ is slightly larger than ${\cal L}_v$, due to the general tendency to have ${\cal E}_x>{\cal E}_v$ ($\tau_x < \tau_v$). The shape (slope) of ${\cal L}_x$ and ${\cal L}_v$ is seen to change over the time; in particular, during dithering, ${\cal L}_x$ and ${\cal L}_v$ are almost parallel due to self-regulation between $x$ and $v$. 

\subsection{Joint PDFs}

Figure \ref{fig3} shows the joint PDFs $p(x,v,t)$ in the $(x,v)$ plane. The overall position of the peaks is as expected, based on the $(\langle x\rangle,\langle v\rangle)$ phase portraits in figure \ref{fig1}. Seeing the full structure however reveals striking features, including strongly non-Gaussian features and multiple peaks. It is also of great interest that the final collapse to $x,v\to0$ does {\it not} consist of a simple motion of the peak toward the origin. Instead, comparing times $t=30,40,50$, we see how the original peak remains largely in the same position, and a secondary peak grows and eventually dominates near the origin. Thus, although $\langle x\rangle$ and $\langle v\rangle$ appear to 
decrease to $\langle x \rangle = \langle v \rangle = 0$ continuously in time, the actual PDFs develop this multimodal structure. 
{Similar (albeit opposite) evolution from a unimodal to bimodal PDF was shown in the Ginzburg-Landau phase transition \cite{phasetransition} where the final bimodal PDF was established by the growth of the new two peaks of a bimodal PDF and the decay of the peak of the initial unimodal PDF. The physics behind such behaviour is that a new (stable) attractor gets stronger while the old (unstable) one becomes weaker in stochastic environment.}
Also, note that for a PDF with more than one peak, mean value (standard deviation) fails to capture the mostly likely value (PDF width), calling for the caution in physical interpretation of these quantities.\\

\subsection{Marginal PDFs}
\begin{figure}
\begin{center}
\includegraphics[scale=0.9]{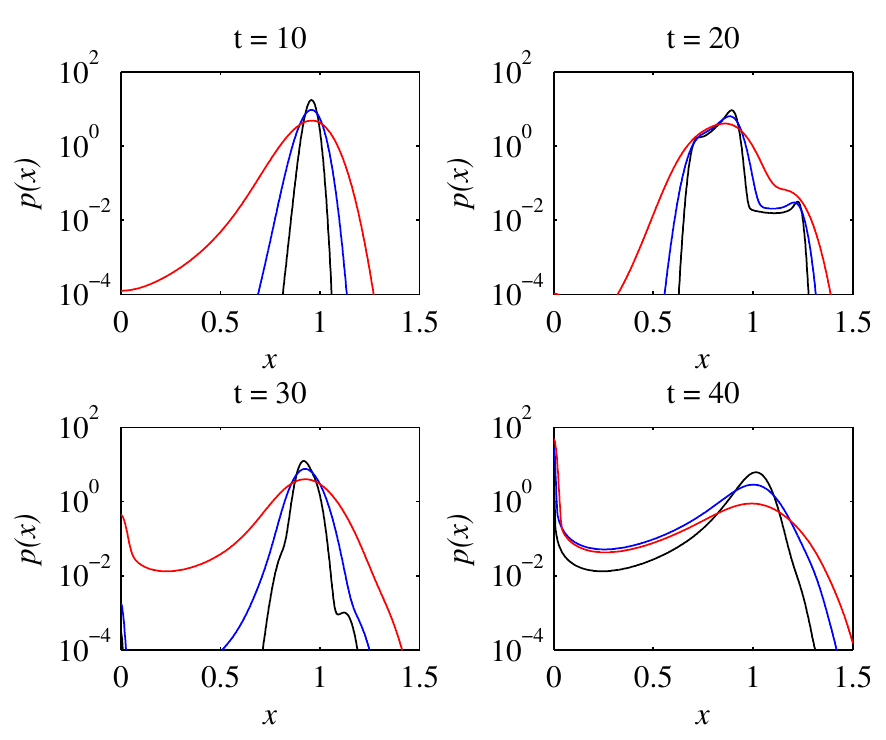}
\vspace{-0.2cm}
\caption{The marginal PDFs $p(x)=\int p(x,v)\,dv$ at $t=10,20,30,40$ as indicated. As in figure 1 the curves are color-coded with [black,blue,red] corresponding to $D_x=[1,4,16]\cdot10^{-4}$ respectively.\\}
\label{fig4}
\end{center}
\end{figure}

\begin{figure}
\begin{center}
\includegraphics[scale=0.9]{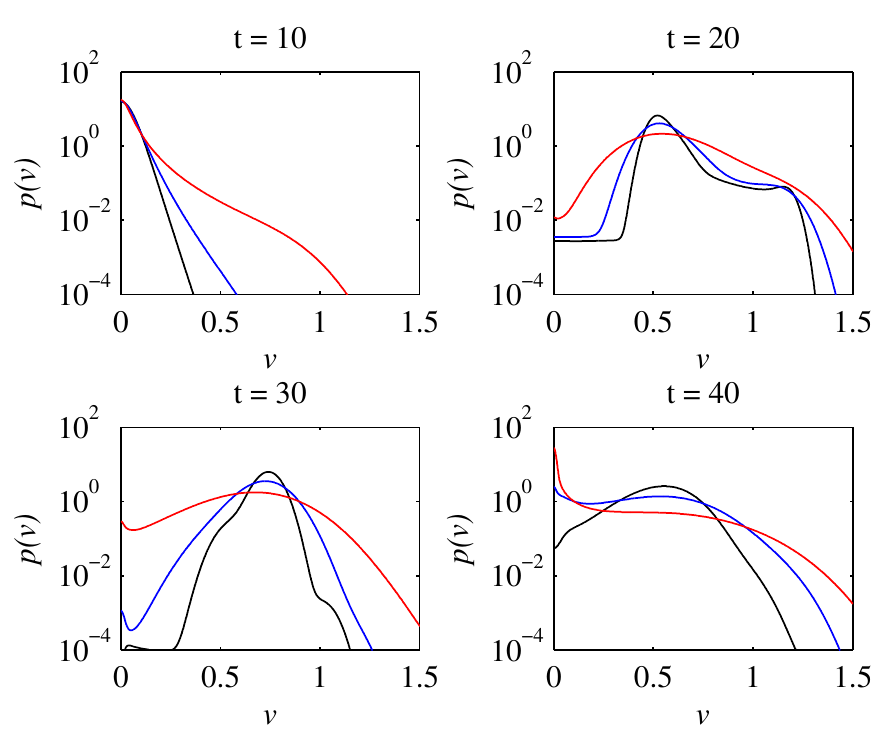}
\vspace{-0.2cm}
\caption{The marginal PDFs $p(v)=\int p(x,v)\,dx$ at $t=10,20,30,40$ as indicated, and with color coding as in figures 1 and 4.\\}
\vspace{0.5cm}
\label{fig5}
\end{center}
\end{figure}

Figures \ref{fig4} and \ref{fig5} finally show the marginal PDFs $p(x,t)$ and $p(v,t)$. We see the strong deviations from Gaussian behavior and a significant asymmetry around the peak even more clearly here than in figure \ref{fig5}, and again the bimodal nature of the L-H transition. Since figures \ref{fig3}-\ref{fig5} are shown only for $x,v\ge 0$, PDFs have multiple peaks in $x,v = (-\infty, \infty)$. 
{Of particular note is the observation that in figures \ref{fig4} and \ref{fig5}, $p(v,t)$ is more stretched than $p(x,t)$ at the right tail; that is, rare events of large $v$ are more common than rare events of large $x$, even though the stochasticity $D_x$ is directly acting on $x$. This effect of $D_x$ to elevate the right tail of $p(v,t)$ more than that of $p(x,t)$ suggests that the transitions to I-phase and H-mode are facilitated by rare events of strong zonal~flow~$v$.}


\section{Conclusions} 
{Our work was motivated by the fact that L-H transition experiments are very expensive, requiring careful planning. In particular, it is desirable that experiments are done in a way to be able to measure the most important quantities. To this end, we have proposed methods based on time-dependent PDFs and information diagnostics that are very sensitive to the dynamics during the L-H transition in terms of elucidating correlation/self-regulation among different players and spatial locations, forecasting, etc.  
While rare, large-amplitude events (e.g.\ blobs) have been thought to be important for enhancing transport, our results for the first time point out the interesting possibility that rare, large amplitude events of strong zonal flow shearing can also play an important role in helping the L-H transition. This provides a new interesting paradigm to be tested {in future works, e.g. by measuring PDFs of zonal flow as well as turbulence in the L-H transition in experiments}.

Practically, to apply our method to experimental data, time-dependent PDFs can be calculated by sampling the data in the time-series of different variables (fluctuating density, electric field, etc.) by using moving-time windows, as was done in a Hasagawa-Wakatani turbulence model \cite{HM2019} where 
information length was shown to be a novel methodology of assessing
the effects of coherent structures and turbulent dynamics in plasmas,
e.g., quantifying the decorrelation of the flux between different spatial
positions due to coherent structures.
{
Therefore, one promising future work will be to utilise our method to predict undesirable plasmas events (e.g.\ ELMs, eruptions) well before other methods can, so that the occurrence of such events can be avoided or else controlled to some degree.}
It will also be of great interest to apply this methodology to understand the temporal-spatial dynamics in other L-H transition turbulence models as well as experimental data to quantify correlations at different spatial positions \cite{HM2019,KIM20}. \\

{\bf Acknowledgement} We thank Yasmin Andrew and DIII-D team members (e.g.\ L. Schmitz, C.F. Maggi, and G. Mckee) for useful discussions. EK acknowledges the Leverhulme Trust Research Fellowship (RF-2018-142-9).

\end{document}